\documentclass[12pt,preprint]{aastex}

\shortauthors{Liu et al.}

\begin{document}

\title{On the Generation and Evolution of Switchbacks and the Morphology of Alfv\'enic Transition: Low Mach-number Boundary Layers}

\author{Ying D. Liu\altaffilmark{1,2}, Hao Ran\altaffilmark{1,2}, Huidong Hu\altaffilmark{1}, and Stuart D. Bale\altaffilmark{3}}
  
\altaffiltext{1}{State Key Laboratory of Space Weather, National Space 
Science Center, Chinese Academy of Sciences, Beijing, China; liuxying@swl.ac.cn}

\altaffiltext{2}{University of Chinese Academy of Sciences, Beijing, China}

\altaffiltext{3}{Space Sciences Laboratory, University of California, Berkeley, CA 94720, USA}

\begin{abstract}

We investigate the generation and evolution of switchbacks (SBs), the nature of the sub-Alfv\'enic wind observed by Parker Solar Probe (PSP), and the morphology of the Alfv\'enic transition, all of which are key issues in solar wind research. First we highlight a special structure in the pristine solar wind, termed a low Mach-number boundary layer (LMBL). An increased Alfv\'en radius and suppressed SBs are observed within an LMBL. A probable source on the Sun for an LMBL is the peripheral region inside a coronal hole with rapidly diverging open fields. The sub-Alfv\'enic wind detected by PSP is an LMBL flow by nature. The similar origin and similar properties of the sub-Alfv\'enic intervals favor a wrinkled surface for the morphology of the Alfv\'enic transition. We find that a larger deflection angle tends to be associated with a higher Alfv\'en Mach number. The magnetic deflections have an origin well below the Alfv\'en critical point, and deflection angles larger than $90^{\circ}$ seem to occur only when $M_{\rm A} \gtrsim 2$. The velocity enhancement in units of the local Alfv\'en speed generally increases with the deflection angle, which is explained by a simple model. A nonlinearly evolved, saturated state is revealed for SBs, where the local Alfv\'en speed is roughly an upper bound for the velocity enhancement. In the context of these results, the most promising theory on the origin of SBs is the model of expanding waves and turbulence, and the patchy distribution of SBs is attributed to modulation by reductions in the Alfv\'en Mach number. Finally, a picture on the generation and evolution of SBs is created based on the results.    

\end{abstract}


\section{Introduction}

The Parker Solar Probe (PSP) mission, launched in 2018 August, is intended to dive below the Alfv\'en surface and make direct in situ measurements of the source regions of the solar wind for the first time \citep{fox2016}. One of the striking features in the observations from the very first encounter is the prevalence of Alfv\'enic flows with reversed magnetic field and enhanced radial velocity, called switchbacks \citep[SBs;][]{bale2019, kasper2019}. Similar structures have been observed in the solar wind by earlier missions but with much reduced frequency \citep[e.g.,][]{balogh1999, gosling2009, horbury2018}. SBs tend to occur in patches, i.e., bursts of the structure separated by quieter intervals \citep[e.g.,][]{bale2021, fargette2021}, so they appear to be modulated. The origin of SBs is highly debated. Models have been proposed to explain the generation and evolution of SBs, including interchange magnetic reconnection \citep[e.g.,][]{fisk2020, zank2020, drake2021}, velocity shear and footpoint motion \citep{landi2006, ruffolo2020, schwadron2021}, expanding waves and turbulence \citep[e.g.,][]{squire2020, shoda2021}, and coronal jets \citep[e.g.,][]{sterling2020}. No sufficient evidence has been provided so far to distinguish the theories.   

The Alfv\'en surface marks the critical transition in the solar wind radial velocity from sub-Alfv\'enic to super-Alfv\'enic \citep{weber1967}. The topological nature of this transition is a key issue but remains unclear. \citet{liu2021} estimate the average Alfv\'en radius to be around 10 solar radii from PSP measurements at the first four encounters. They observe a speed dependence in the Alfv\'en radius, which suggests a ``rugged" Alfv\'en surface around the Sun. Similar results are obtained by \citet{verscharen2021}. Based on MHD simulations coupled to a turbulent transport model, \citet{chhiber2022} argue that the transition may be better described as an extended, fragmented zone rather than a wrinkled surface, i.e., a region with mixed parcels of sub-Alfv\'enic and super-Alfv\'enic wind. Again, evidence is needed to distinguish this hypothesis from that of a wrinkled surface. 

As another accomplishment, PSP entered a sub-Alfv\'enic solar wind region for about 5 hr around a distance of about 19 solar radii at the eighth encounter on 2021 April 28 \citep{kasper2021}. This first glimpse of the sub-Alfv\'enic wind implies a surprisingly large Alfv\'en radius, which is over 20 solar radii. However, the solar wind density within all the first several sub-Alfv\'enic intervals is unusually low, which is a main factor contributing to the crossings of the Alfv\'enic transition \citep{kasper2021}. This indicates that the observed sub-Alfv\'enic wind is not a typical wind. At the subsequent encounters, crossings of the Alfv\'enic transition have been made on several more occasions \citep[e.g.,][]{bandyopadhyay2022}. Measurements in those sub-Alfv\'enic intervals show predominantly outward propagating Alfv\'enic fluctuations \citep[e.g.,][]{zank2022} and much weaker, fewer SBs than in the super-Alfv\'enic wind \citep[e.g.,][]{kasper2021, bandyopadhyay2022}. The nature of the special, sub-Alfv\'enic wind remains to be investigated, and an intriguing question is what it reveals concerning the morphology of the Alfv\'enic transition and the generation of SBs.

This paper has a threefold focus. First, we identify a special structure in the pristine solar wind, defined as a low Mach-number boundary layer (LMBL), and determine its typical signatures and source region on the Sun. Second, we illustrate that the sub-Alfv\'enic wind observed by PSP is an LMBL flow by nature, and the similar origin and similar properties of the sub-Alfv\'enic LMBLs suggest a wrinkled surface for the morphology of the Alfv\'enic transition. Third, we show how the properties of SBs and those of the LMBLs themselves help clarify the generation and evolution of SBs as well as their modulation. Data and methodology are described in Section 2. Details on LMBLs and their interpretations are given in Section 3. We then discuss in Section 4 important implications on the morphology and crossings of the Alfv\'enic transition, the generation and evolution of SBs, and the SB modulation. The conclusions are summarized in Section 5. 

\section{Data, Methodology and Speculations}

We use measurements from the FIELDS instrument suite \citep{bale2016} and the SWEAP package \citep{kasper2016} aboard PSP. The SWEAP ion instruments include an electrostatic analyzer (SPAN-I) on the ram side of PSP, and a Faraday Cup (SPC) that looks directly at the Sun \citep{kasper2016, case2020, whittlesey2020}. With their combined fields of view, SPAN-I and SPC are able to provide plasma parameters of the solar wind throughout key encounter phases of the mission. Parameters of the alpha particles are also available starting from the fourth encounter, as PSP moves fast enough to shift the majority of the particle velocity distribution core into the field of view of SPAN-I. Because of the partial moments in the plasma measurements, the electron density from quasi-thermal noise (QTN) spectroscopy \citep{moncuquet2020} is used as a proxy of the plasma density. This density is considered to be most reliable, since it is derived from measurements of the local plasma frequency. All the solar wind parameters from FIELDS and SWEAP are interpolated to 2-s, evenly spaced averages. 

\subsection{Derivation of Deflection Angle}

The deflection angle of the magnetic field is defined as the deviation angle from the average background field direction, which is assumed to be radial or anti-radial depending on the field polarity. This assumption is valid at PSP encounter distances, because at those distances the Parker spiral field is indeed almost radial or anti-radial. In RTN coordinates, for an overall positive field (i.e., $\langle{B_R}\rangle > 0$) the deflection angle can be expresses as        
\begin{equation}
\theta = \left\lbrace \begin{array}{ll}
\arccos(B_R/B),~~~~~~~~~~~~~~~~~B_T>0, \\
-\arccos(B_R/B),~~~~~~~~~~~~~~B_T<0.
\end{array} \right.
\end{equation}
For an overall negative field (i.e., $\langle{B_R}\rangle < 0$) the deflection angle is written as 
\begin{equation}
\theta = \left\lbrace \begin{array}{ll}
-\arccos(-B_R/B),~~~~~~~~~~~~~B_T>0, \\
\arccos(-B_R/B),~~~~~~~~~~~~~~~~B_T<0.
\end{array} \right.
\end{equation}
An advantage of such a definition is that the sign of $\theta$ indicates the deflection direction. A positive angle ($0<\theta\leqslant\pi$) corresponds to a counterclockwise deflection in the RT plane as viewed from the north, while a negative one ($-\pi\leqslant\theta<0$) a clockwise deflection. It takes a value of $\pi$ or $-\pi$ when the magnetic field is completely reversed. This definition also allows to set up a relationship with the enhancement in the solar wind radial velocity, as will be shown below. Note that our definition should not be confused with the magnetic field cone angle, the angle between the local magnetic field vector and the radial direction \citep[e.g.,][]{horbury2018, kasper2019, woolley2020} that does not provide information on the deflection direction. 

An SB by name implies a deflection angle larger than $90^{\circ}$. To look at how the phenomenon forms and evolves, however, we need to include all possible deflection angles, not just those more than $90^{\circ}$ \citep[e.g.,][]{mozer2021, tenerani2021}. Indeed, \citet{wit2020} find a continuum of deflections with no particular threshold that can be used to define an SB. As will be shown below, those small deflections hold important clues on the origin of the phenomenon in the solar atmosphere (below the Alfv\'en critical point), while those large ones indicate key information on more evolved states of the phenomenon at greater distances. We suggest that the terminology is better changed to Alfv\'enic flows with deflected magnetic field and enhanced radial velocity, or Alfv\'enic deflections for short. Section~2.2 provides more support for this statement.   

\subsection{Velocity Enhancement and Relationship with Deflection Angle} 

For a magnetic field with magnitude $B$ deflected from its original direction (radial or anti-radial depending on the field polarity) by an angle $\theta$, the change in the radial component is $\delta{B_R} = B_{R0}\cos\theta - B_{R0}$, where $B_{R0}$ is the initial radial component of the field. Obviously, $B=|B_{R0}|$ under our assumption. For Alfv\'enic fluctuations, the change in the radial velocity is related to the change in the radial field by $\delta{v_R}/v_{\rm A} = \pm \delta{B_R}/B$ \citep[e.g.,][]{belcher1971}, where $v_{\rm A}$ is the local Alfv\'en speed. Outward propagating Alfv\'enic fluctuations yield a negative sign for the above equation if the background field is positive, and a positive sign if the background field is negative. It can be readily shown that
\begin{equation}
{\delta{v_R} \over v_{\rm A}} = 1 - \cos\theta 
\end{equation}
for outward propagating Alfv\'enic fluctuations. This simple analytical development indicates that any deflection of the field would always be associated with an enhancement in the solar wind radial velocity (i.e., $\delta{v_R}>0$). It explains the one-sided nature of the velocity spikes observed within SBs \citep[e.g.,][]{gosling2009, horbury2018, kasper2019}, which is simply a result of outward propagating Alfv\'enic fluctuations. At PSP encounter distances the mean field is almost radial or anti-radial, so the effect is amplified in the encounter measurements compared with those at greater distances where the field has larger spiral angles. This predicts that considerable radial velocity spikes in SBs observed at 1 AU and beyond may correspond to a background field geometry with a large radial component. 

Another interesting prediction is that below the Alfv\'en critical point the velocity enhancement and thus the deflection angle must be significantly reduced. This can be shown with the following simple reasoning. Equation~(3) indicates that the velocity is always enhanced and the enhancement can be a large fraction of the local Alfv\'en speed. If it were not reduced below the Alfv\'en critical point, the enhancement would easily lead to a radial velocity higher than the local Alfv\'en speed, which conflicts with the condition of $v_R < v_{\rm A}$. A prerequisite is that Alfv\'enicity still holds for the sub-Alfv\'enic wind, which is indeed the case \citep{zank2022}.    

One may expect from Equation~(3) an enhancement of $v_{\rm A}$ in the radial velocity for a $90^{\circ}$ deflection and 2$v_{\rm A}$ for a complete field reversal. Note that, however, the background magnetic field at PSP encounter distances is not exactly radial or anti-radial (i.e., $B \geqslant |B_{R0}|$), so the equation is an upper limit for the velocity enhancement. To quantify the velocity enhancement and how it compares with Equation~(3), we derive the velocity enhancement by applying a low-pass Butterworth filter to the observed radial velocities with a cutoff frequency of 2$\times$$10^{-4}$ Hz. This cutoff frequency is well below the break frequency (about 2$\times$$10^{-3}$ Hz) separating the energy-containing range and inertial range in the power spectrum of solar wind fluctuations \citep[e.g.,][]{kasper2021, zank2022}. Velocity variations with frequencies higher than the cutoff frequency are removed in the filtering. The radial velocity enhancement can then be calculated using $\delta{v_R} = v_R - v_{Rf}$, where $v_{Rf}$ is the filtered value and serves as the ``baseline" for the solar wind radial velocity.   

When writing this paper, we notice a similar model proposed by \citet{matteini2014} to explain the correlation between the solar wind speed and the local magnetic field orientation in fast, polar coronal hole wind observed by Ulysses. They also suggest that SBs may naturally arise from Alfv\'enic fluctuations in the context of their model and the observed correlation. Our independent development of Equation~(3) closely parallels their work. New important information will be revealed by looking at how the velocity enhancement varies with the Alfv\'en Mach number as well as the deflection angle. These may constitute a more complete explanation for the generation and evolution of SBs.   

\subsection{Estimate of Alfv\'en Radius}

The derivation of the Alfv\'en radius ($r_{\rm A}$) and its uncertainties are described in \citet{liu2021}. Here we rewrite the expression in RTN coordinates 
\begin{equation}
r_{\rm A} = \sqrt{v_R \over v_{R\rm A}}{r \over M_{\rm A}},
\end{equation}
where $r$ is the heliocentric distance, $v_{R\rm A}$ the radial velocity at the Alfv\'en critical point (i.e., $r_{\rm A}$), and $M_{\rm A}$ the radial Alfv\'en Mach number. In the classic theory of \citet{weber1967}, the radial Alfv\'en Mach number is defined as $M_{\rm A} = v_R\sqrt{\mu\rho}/|B_R|$, where $\rho$ is the mass density of the solar wind and $\mu$ the permeability constant. We realize that SBs may give rise to a singularity for a deflection around $90^{\circ}$, since $|B_R|=B\cos\theta$. Indeed, we see large spikes in $M_{\rm A}$ \citep{liu2021}. While this could be useful for identifying SBs, we redefine it as $M_{\rm A} = v_R\sqrt{\mu\rho}/B$ taking advantage of $B \simeq |B_R|$ in the background solar wind at PSP encounter distances.       

A reasonable approximation is obtained by assuming that $v_R$ does not change much from $r_{\rm A}$ to PSP encounter distances, i.e.,    
\begin{equation}
r_{\rm A} \simeq {r \over M_{\rm A}}.
\end{equation}
\citet{weber1967} demonstrate that this approximation, while simple, provides a rigorous estimate of the Alfv\'en radius. This is particularly true for PSP encounter measurements as they are made close to the Sun. The estimate should be considered as a lower limit when PSP is outside $r_{\rm A}$ (i.e., $v_R > v_{R\rm A}$), and an upper limit when inside (i.e., $v_R < v_{R\rm A}$). Readers are directed to \citet{liu2021} for more detailed discussions on the uncertainties of the estimate.     

\subsection{Determination of Solar Source} 

The magnetic connectivity between PSP and the photosphere is established with a potential field source surface (PFSS) model plus a simple ballistic projection from the location of the spacecraft to the source surface \citep[e.g.,][]{altschuler1969, schatten1969, wang1992, badman2020}. An ideal Parker spiral is assumed for the magnetic field line connecting PSP to the source surface with curvature determined by the solar wind radial velocity observed at PSP. Below the source surface the coronal magnetic fields are constructed using the PFSS model, based on the Air Force Data Assimilative Photospheric Flux Transport (ADAPT) magnetograms provided by the Global Oscillation Network Group (GONG). The ADAPT-GONG synoptic map is updated every two hours. We select a map when the PFSS model becomes stable with time. For encounter 2, which is included in our analysis, the PFSS model results vary significantly between the ADAPT-GONG maps because of the emergence of an active region on the far side of the Sun \citep{wallace2022}. The PFSS model for encounter 2 is thus developed using a magnetogram recorded on 2019 April 12 when the far-side active region is incorporated into the map. 

The height of the source surface is usually set to 2.5 solar radii ($R_S$), at which the field lines are forced to be radial. The PFSS modeling gives areas of open fields, which can be compared with EUV imaging observations of coronal holes. This is how we select a value for the height of the source surface. In this study we use EUV synoptic maps from SDO 193 \AA\ observations for such a comparison (available at https://solarflare.njit.edu). The magnetic mapping from the PSP location to the source surface and then to the photosphere suffers different sources of uncertainties, including the radial variation of the solar wind speed and the magnetogram used as the boundary condition. The uncertainties are difficult to quantify, but some of the effects may counteract each other \citep[see more discussions in][]{badman2020, kasper2021, pinto2021}. Our experience with the mapping indicates that the position error on the photosphere is at least a few degrees \citep{chen2021, meng2022}.   

\section{Examples of LMBLs and Interpretations}

Figure~1 shows the in situ measurements near perihelion of encounter 2, where a transition from slow to relatively fast wind is observed. We see a clear drop in the radial Alfv\'en Mach number associated with this transition. The boundary of the LMBL is mainly determined from the decrease in $M_{\rm A}$. The primary contribution to this drop is the significantly reduced density plus the relatively low speed. The Alfv\'en radius increases to more than 20 solar radii. If the altitude of PSP were as low as 20 $R_S$, we would see a crossing of the Alfv\'enic transition as early as encounter 2. \citet{liu2021} indicate conditions of reduced densities and low radial velocities that favor crossings of the Alfv\'enic transition. The crossings starting from encounter 8 confirm this prediction \citep[][and see data below]{kasper2021}. Note that we scale the density and magnetic field to values at 1 AU in order to eliminate their radial variations associated with the change in the distance of the spacecraft. The magnetic field strength is roughly constant across the slow-to-fast wind transition. \citet{rouillard2020} examine the same time period and suggest that PSP was measuring high-density streamer flows when it was connected to streamers and tenuous wind when it was not. The low density and steadily increasing velocity inside the LMBL indicate that PSP was likely sampling open magnetic fields from a coronal hole near its boundary, and the even higher velocity and persistent low density after the LMBL are suggestive of deeper penetration into the coronal hole wind. Our magnetic mapping will provide further evidence for this interpretation. 

An important thing to note is that the amplitudes of SBs within the LMBL are considerably reduced. A first impression of this reduction is from the variations of $B_{R}$ (Figure~1d). Quantitative results about the reduction are given by the deflection angle of the magnetic field and its hourly rate of over $30^{\circ}$. The solar wind radial velocity exhibits largely one-sided variations (i.e., enhancement) compared with its ``baseline" from the low-pass filtering. Inside the LMBL, the velocity enhancement in Figure~1c does not show an obvious decrease, but when it is calculated in units of the local Alfv\'en speed a clear reduction is seen (Figure~1h). These are consistent with the correlation between the deflection angle and velocity enhancement predicted by Equation~(3). In general, $\delta{v_R} \leqslant v_{\rm A}$ as indicated by Figure~1h (see more discussions in Section~4). Also note that the LMBL has fine structures: whenever $M_{\rm A}$ drops we see a reduction in SBs. This seems true for data outside the LMBL as well. On average, the slower wind ahead of the LMBL has larger deflection angles than the faster wind behind the LMBL, which is probably due to the higher $M_{\rm A}$ of the slower wind. Therefore, SBs are better sorted by $M_{\rm A}$ rather than the velocity \citep[also see][]{mozer2021}. The deflection angle before the LMBL is predominantly negative, indicative of a preferential clockwise rotation in the magnetic field associated with the streamer flows. This agrees with the finding of \citet{meng2022}. 

Figure~2 presents the in situ measurements near perihelion of encounter 8. Compared with the case in Figure~1, an opposite transition is observed, i.e., from relatively fast, tenuous to slow, dense wind. PSP was likely transiting from a deeper immersion in a coronal hole wind before the LMBL, to the stream out of the coronal hole boundary in the LMBL, and then to streamer flows after the LMBL. Indeed, we see crossings of the heliospheric current sheet (HCS) embedded in the high-density wind. Significant decreases in the Alfv\'en radius are observed around the HCS crossings, which is predicted by MHD simulations \citep{chhiber2022}. The alpha particles show a radial velocity higher than that of the protons for most of the wind inside the LMBL. This is a typical signature of fast wind \citep[e.g.,][]{asbridge1976, marsch1982}, although the solar wind is slow in our case. The characteristics of being slow while resembling fast wind provide support for our interpretation about the origin of an LMBL, i.e., boundary wind from inside a coronal hole.  

The situation is in general akin to what is shown in Figure~1, such as associations of the LMBL with decreased $M_{\rm A}$, increased $r_{\rm A}$ and reduced SBs. In the current case, the Alfv\'en radius rises above the distance of PSP, which leads to the first glimpse of the sub-Alfv\'enic wind \citep{kasper2021}. The fact that an LMBL has a large Alfv\'en radius enabling an easier crossing of the Alfv\'enic transition suggests that the sub-Alfv\'enic wind observed by PSP so far is an LMBL flow by nature. More specifically, the sub-Alfv\'enic wind is a special type of wind flowing out along open magnetic fields from a coronal hole near its boundary. Note that in the sub-Alfv\'enic wind SBs (or Alfv\'enic deflections in our terminology) do not disappear; they are just reduced because of the low $M_{\rm A}$. This observation is consistent with our speculation about the behavior of SBs below the Alfv\'en critical point based on Equation~(3). 

Figure~3 displays the in situ measurements near perihelion of encounter 9. Again, we see a transition from relatively fast, tenuous to slow, dense wind. During the transition in the tenuous wind, PSP was likely penetrating deeper inside a coronal hole wind (i.e., out of the boundary wind) temporarily. This is why two LMBLs are identified and relatively fast wind with an SB patch is seen between the two LMBLs. The solar wind velocity is also enhanced behind the second LMBL. This is not a coronal hole wind, but reconnection outflow around the HCS crossing \citep[e.g.,][]{phan2020, chen2021}. Inside the two LMBLs, the radial velocity of the alpha particles is considerably higher than that of the protons, despite the fact that the solar wind is slow. This, once more, supports our interpretation about the origin and nature of an LMBL.  

Like the picture that we have seen in Figures~1 and 2, the two LMBLs are associated with decreased $M_{\rm A}$, increased $r_{\rm A}$ and reduced SBs. Again, the sub-Alfv\'enic wind in the present case is an LMBL flow by nature. Interestingly, here we can compare the case when PSP was ``surfing" around the Alfv\'enic transition (i.e., $M_{\rm A} \simeq 1$ inside LMBL1) with the case when PSP was ``diving" well below it (i.e., $M_{\rm A} < 1$ inside LMBL2). The SBs (better called Alfv\'enic deflections) within LMBL2 are further reduced compared with those inside LMBL1 (Figure~3d, h). The magnetic field in LMBL2, however, still shows an indication of deflections (Figure~3d), so the magnetic deflections have an origin well below the Alfv\'en critical point. They develop when $M_{\rm A}$ increases or the plasma accelerates. At a certain $M_{\rm A}$ or distance, their amplitudes are such that the magnetic field begins to be deflected backward. The occurrence of $90^{\circ}$ deflections should be well above the Alfv\'en critical point, as suggested by our results. 

More evidence for the above interpretations is provided by the magnetic mapping of the LMBLs, as shown in Figure~4. The LMBL from encounter 2 is connected to the boundary of the equatorial extension of a southern polar coronal hole (Figure~4a). A similar mapping result is obtained by \citet{meng2022} and \citet{wallace2022} for the same time period. This magnetic connectivity agrees with our interpretation of the LMBL as the transition layer between streamer flows and deeper interior of a coronal hole wind. For the LMBL from encounter 8, the magnetic mapping shows connections first to the edge of a low-latitude small coronal hole, then to the boundary of the equatorial extension of a southern polar coronal hole, and finally to the edge of another low-latitude small coronal hole (Figure~4b). The mapping result is quite similar to that in \citet{kasper2021}. Although there are multiple coronal holes in the present case, the connections to their boundaries are consistent with our interpretation of the LMBL being coronal hole boundary wind. It is important to note that the first low-latitude coronal hole is located near a small active region (see the background EUV image in Figure~4b). The magnetic field is enhanced at the footpoints of the field lines, which may result in a higher rate of energy deposition into the wind. This may explain the large fluctuations in the solar wind radial velocity, in particular the radial velocity of the alphas (Figure~2c). As for the LMBLs from encounter 9, connections to the boundary of a low-latitude coronal hole are seen once more (Figure~4c). We have indicated earlier that the interval between the two LMBLs is connected to deeper interior of a coronal hole (red lines in Figure~4c). How deep this is cannot be resolved by the magnetic mapping given its uncertainties.  

Figure~5 illustrates the origin and nature of an LMBL, based on information collected from PSP measurements and the magnetic mapping. The solar wind constituting the LMBL emanates from the peripheral region inside a coronal hole along rapidly diverging open field lines. The origin from within the coronal hole explains the observed low density and considerable differential velocity between alphas and protons. It is well known that there is an inverse relationship between the solar wind speed and coronal field line expansion factor \citep[e.g.,][]{wang1990}. The rapidly diverging field lines have a large expansion factor, which explains the low velocity of the wind. \citet{wang2000} suggest that the area just inside the coronal hole boundary is a source of slow wind. Although slow, the wind may bear signatures of fast wind given its origin. Slowly diverging open field lines are rooted deeper into the coronal hole. They have a smaller expansion factor, so the wind speed is higher. The LMBL can thus be considered as a transition layer between slow and fast wind. The width of the transition layer is expected to be a few degrees on the Sun \citep[also see][]{zurbuchen1999}, i.e., comparable to or smaller than the uncertainty of the magnetic mapping. Although this is a small region on the Sun, it can map to a considerable volume in the heliosphere. 

This simple picture clarifies many observations associated with the LMBLs and sub-Alfv\'enic wind. The decrease in the radial Alfv\'en Mach number is a direct consequence of the reduced density and relatively low velocity, which in turn leads to the increase in the Alfv\'en radius and the suppression of SBs. The enhanced Alfv\'en radius allows an easier crossing of the Alfv\'enic transition, so the sub-Alfv\'enic wind is observable even at a distance of about 20 $R_S$ from the Sun. The substantial volume that the LMBL maps to in the inner heliosphere implies that detection of the sub-Alfv\'enic wind is not occasional. An important thing to keep in mind though is that the sub-Alfv\'enic wind is not a typical wind, but a special wind emanating from a coronal hole near its boundary along rapidly diverging open field lines. The low Mach number will inhibit the amplitudes of SBs given the Alfv\'enicity in the velocity and magnetic field fluctuations, as predicted by Equation~(3). Note that there may be other sources on the Sun for LMBLs, such as patches of rapidly diverging open fields near an active region or from the entire interior of a very small coronal hole; the field line geometry may also lead to a reduced density and relatively low speed.   

We summarize typical signatures of an LMBL that can be used as criteria to identify the phenomenon from solar wind measurements. The primary signatures include a decreased radial Alfv\'en Mach number, a reduced density, a relatively low velocity, and suppressed SBs. Although the decreased Alfv\'en Mach number is a result of the reduced density and relatively low velocity, it is always helpful to look at all the three parameters together. Suppression in the amplitudes of SBs is useful in the identification of the boundaries of an LMBL. Specific thresholds for these parameters are difficult to quantify because of variations of the quantities with distance. The timescale of an LMBL also varies depending on how long the magnetic footpoint of the spacecraft lingers at the source and the size of the source on the Sun.

\section{Important Implications}

\subsection{Morphology and Crossings of Alfv\'enic Transition}

A first implication concerns the morphology and crossings of the Alfv\'enic transition. \citet{chhiber2022} argue that fluctuations in the plasma and magnetic field, which they expect to increase toward the Alfv\'enic transition, would lead to enhanced variations in the Alfv\'en Mach number approaching the transition. Based on this reasoning, they suggest that the Alfv\'enic transition should not be a simple surface (either smooth or corrugated), but an extended, fragmented zone with mixed parcels of sub-Alfv\'enic and super-Alfv\'enic wind. They demonstrate this possibility using MHD simulations coupled to a turbulent transport model. The possibility is fascinating in view of the turbulent nature of the corona and solar wind. As a result of their theory, one would see subvolumes or blobs of sub-Alfv\'enic wind of random origin and random character in regions that are predominantly super-Alfv\'enic. Our results, however, indicate that the LMBLs including the sub-Alfv\'enic wind all have a similar origin (i.e., coronal hole boundaries with rapidly diverging open fields) and similar properties (i.e., reduced density, relatively low velocity, and signatures resembling fast wind). These are difficult to explain in the frame of their theory, but favor the picture of a ``rugged" Alfv\'en surface that \citet{liu2021} have proposed earlier. 

The typical or average Alfv\'en radius is 10 - 12 solar radii. This value is slightly enhanced by the modification of the radial Alfv\'en Mach number, compared with that of \citet{liu2021}. With this average Alfv\'en radius, we echo the suggestion of \citet{liu2021} that substantial diving below the Alfv\'en surface is plausible only for relatively slow solar wind given the orbital design of PSP. The sub-Alfv\'enic wind observed by PSP is, again, not a typical wind. Its detection is simply a result of an enhanced Alfv\'en radius that enables an easier crossing of the Alfv\'enic transition. We do anticipate higher frequency of sub-Alfv\'enic intervals as PSP descends to lower perihelia. For example, flows with magnetic footpoints located deeper inside a coronal hole will have a higher possibility that their Alfv\'en radius is above PSP distance, as the spacecraft descends. This can be easily seen in the context of our results.    

\subsection{Generation and Evolution of Switchbacks}

Another implication pertains to the generation and evolution of SBs, a highly debated issue in the field. Our results indicate a dependence of the amplitudes of SBs on the radial Alfv\'en Mach number. Figure~6 shows the magnetic field deflection angle versus the Alfv\'en Mach number using encounter 9 measurements as a representative. The measurements were made from 15:00 UT on August 5 to the end of August 9, 2021 when the distance of PSP ranged from about 16 to 45 $R_S$. This time period is chosen mainly because it covers a relatively broad spectrum of $M_{\rm A}$; during the time period, there are no HCS crossings or transient phenomena such as coronal mass ejections.    

The distribution resembles a ``herringbone" structure, where a larger deflection angle tends to be associated with a higher $M_{\rm A}$ (Figure~6 left). Note that the trend with $M_{\rm A}$ is contaminated by mixing of different solar wind streams of different origins or properties. Striations are discernible in the distribution, which are likely formed by flows of similar origin or properties. The increasing trend of the deflection angle with $M_{\rm A}$ is better seen along a striation. It is most clear in the sub-Alfv\'enic wind (within the dashed blue box) where mixing of different flows is reduced. If the increase in $M_{\rm A}$ is considered as the solar wind acceleration process from low to high altitudes, then we would expect larger deflection angles at greater distances. This is indeed found by \citet{mozer2020}. Deflection angles larger than $90^{\circ}$ seem to occur only when $M_{\rm A} \gtrsim 2$. Obviously, a true SB takes place well above the Alfv\'en critical point. For the majority of data the deflection angles are below $90^{\circ}$, so most magnetic field lines are not deflected backward but just ``sideways". The distribution is not symmetric around the $\theta=0$ line; more negative angles are observed. Specifically, the data show a preferential clockwise deflection in the RT plane. This may be caused by the Parker spiral geometry of the mean field, which rotates clockwise.    

Effects associated with the original definition of $M_{\rm A}$ are shown in the right panel of Figure~6. Similar features are present as in the left panel, including striations in the distribution, increasing trend of $\theta$, predominance of deflections less than $90^{\circ}$, and preferential clockwise deflections. Large values are seen in the $M_{\rm A}$ as a result of the deflection of the magnetic field from the radial or anti-radial direction. A singularity occurs at $90^{\circ}$, which separates the deflections of $\theta > 90^{\circ}$ from the rest. Note that the apparent decreasing trend in the deflection angles larger than $90^{\circ}$ is an illusion caused by the mathematical singularity. In the current case we also see a critical $M_{\rm A}$ value above which deflections larger than $90^{\circ}$ occur, although the value is changed. 

Figure~7 displays the radial velocity enhancement in units of the local Alfv\'en speed versus the magnetic field deflection angle using data from the same time period as in Figure~6. The velocity enhancement in units of the local Alfv\'en speed generally increases with the deflection angle. This is what we expect from Equation~(3). As mentioned earlier, the red curve represents an upper limit of the data, because the mean magnetic field at PSP encounter distances is not exactly radial or anti-radial. Note that, while there are quite some deflection angles larger than $90^{\circ}$, the data points with $\delta{v_R} > v_{\rm A}$ are very few. The fact of $\delta{v_R} \leqslant v_{\rm A}$ in general can also be seen in Figures~1-3. This is surprising since previous studies indicate a typical velocity change of the order of 2$v_{\rm A}$ for a complete field reversal \citep[e.g.,][]{horbury2018, kasper2019, woolley2020}. It seems to suggest a nonlinearly evolved, saturated state where the local Alfv\'en speed is roughly an upper bound for the velocity enhancement. This nonlinearly evolved state, which is missed in previous studies, may hold crucial information on the evolution of SBs.   
 
With the results on the distribution of the amplitudes of SBs as well as the LMBL properties, we are now in a position to test the theories proposed to explain the origin of SBs. We will see how they fit, or are challenged by, our results. Note that the following testing should not be considered as conclusive. More studies are needed to further constrain or distinguish the theories.

(i) Interchange reconnection. A propagating kink in the coronal magnetic field can be produced by reconnection between adjacent open and closed fields \citep[e.g.,][]{fisk2020, zank2020, liang2021}. \citet{drake2021} suggest that reconnection between open and closed flux with a strong guide field injects flux ropes, which can maintain their field geometry over long distances and reproduce some features of SBs observed in the solar wind. Interchange reconnection is an intriguing theory, but a major difficulty it faces is that one would expect decreasing deflection angles with distance. Our results actually reveal increasing deflections with $M_{\rm A}$ or distance (in the solar wind acceleration process $M_{\rm A}$ increases with distance). In addition, an LMBL represents the open-closed boundary, forming an ideal environment for interchange reconnection. Strong SBs are anticipated in an LMBL according to the interchange reconnection theory. We observe the opposite.      

(ii) Velocity shear and footpoint motion. The interaction between Alfv\'enic fluctuations and velocity shears is proposed as a possible mechanism to generate SBs \citep[e.g.,][]{landi2006, ruffolo2020}. Shear-driven dynamics are triggered only above the Alfv\'en critical point \citep{ruffolo2020}, so this can be viewed as an in situ formation process in the solar wind. However, we do see magnetic deflections below the Alfv\'en critical point (i.e., $M_{\rm A} < 1$), although the deflection angles are small. Also, we expect enhanced velocity shears inside an LMBL since it is a transition layer between slow and fast wind. Indeed, \citet{pinto2021} find that strong shears develop at such boundaries using MHD simulations. The velocity shear theory would predict enhanced SBs inside an LMBL, but again the opposite is observed. 

\citet{schwadron2021} suggest that footpoint motion from the source of slow to fast wind, the so-called super-Parker spiral case, is able to create SBs via the velocity shear between the slow and fast wind. When footpoint motion is reversed (i.e., from the source of fast to slow wind), the so-called sub-Parker spiral case, the field line is straightened, i.e., no SBs. Our Figures~1 and 2 can be considered as the super-Parker and sub-Parker spiral cases, respectively. However, SBs are reduced in both transition layers (i.e., the LMBLs) due to the low Alfv\'en Mach number. No difference is seen between the two LMBLs. They find paucity of SBs within plasma rarefaction regions and use this as a support for their theory. Our results indicate that this is just a consequence of the decreased $M_{\rm A}$ in the rarefaction regions because of the reduced density.

(iii) Expanding waves and turbulence. MHD simulations indicate that Alfv\'en waves in the expanding solar wind can naturally develop into SBs as observed at PSP \citep[e.g.,][]{squire2020, shoda2021}. This is the most promising theory in the context of our work, i.e., the increasing trend in the amplitudes of SBs with the Alfv\'en Mach number, and the association with velocity enhancement for any deflection of the field as a result of outward propagating Alfv\'enic fluctuations. Although \citet{squire2020} stress that this is an ``in situ" formation process in the solar wind, the theory can be modified to include the growth of magnetic deflections in the solar atmosphere, such as the work by \citet{shoda2021}. Our results suggest that SBs have an origin in the solar atmosphere. Their amplitudes are just suppressed below the Alfv\'en critical point, and those that have deflection angles larger than $90^{\circ}$ represent a well evolved state above the Alfv\'en critical point.

(iv) Coronal jets. Another possibility is that SBs originate from coronal jets \citep[e.g.,][]{sterling2020}. This remains a speculation so far, and no much information can be compared with our results. Our work suggests that, however, outward propagating Alfv\'enic fluctuations naturally produce magnetic deflections as well as velocity spikes observed in SBs. Therefore, it is not clear if coronal jets are necessary for (or how they contribute to) the generation of SBs. The concern is valid, in particular when we consider that convection motions in the photosphere would readily drive Alfv\'en waves filling the corona and heliosphere \citep[][and references therein]{cranmer2009}.     

\subsection{Modulation of Switchbacks} 

Our results about the LMBLs and dependence of SBs on the Alfv\'en Mach number can be used to explain the modulation of SBs. Previously the patchy distribution of SBs has been attributed to modulation by supergranulation on the Sun \citep{bale2021, fargette2021}; the patches are thought to arise from the diverging magnetic funnels associated with supergranulation, as their angular scales are comparable, i.e., a few degrees. \citet{shi2022} examine measurements from encounters 1 and 10 when the longitudinal span of PSP is only several degrees, and also find similar patchy distributions. They argue that those patches are likely not related with the spatial scale of supergranules and temporal effects have to be taken into account. Here we suggest that between adjacent patches SBs are inhibited by a reduction in the Alfv\'en Mach number. More specifically, the patchy distribution is simply a result of modulation by reductions in $M_{\rm A}$.

To test this idea, we show in Figure~8 the data from the same time period as in the Figure~2 of \citet{bale2021}. The shaded regions separating patches that have been studied by \citet{bale2021} are indeed associated with a reduced Alfv\'en Mach number. Clear decreases are seen in the amplitudes of SBs inside the shaded intervals, including the deflection angle and velocity enhancement in units of the local Alfv\'en speed. In general, we observe a correlation between the drop in $M_{\rm A}$ and the reduction in SBs throughout the time period, except the transient structure on September 26. The correlation is also observed in Figures~1-3. The shaded intervals have similar properties to an LMBL, such as a reduced density, velocity valley, and increased Alfv\'en radius. It is likely that the funnels associated with supergranulation may lead to a higher $M_{\rm A}$ of the wind, but this cannot be determined by the magnetic mapping as its uncertainty is comparable to or larger than the angular size of supergranulation. Our results indicate that a repeated in-and-out motion of the PSP footpoint with respect to one or more LMBLs can readily explain the patchy distribution of SBs, no matter whether it is a spatial or temporal effect.  

\section{Conclusions}

We have highlighted a special structure in the young solar wind, termed a low Mach-number boundary layer (LMBL). Key findings are revealed concerning the nature of the sub-Alfv\'enic wind observed by PSP, the morphology of the Alfv\'enic transition, and the generation and evolution of SBs. We summarize the results as follows.  
  
(1) An LMBL, by name, is characterized by a reduced radial Alfv\'en Mach number resulting from its decreased solar wind density and relatively low velocity. The low Mach number, in turn, leads to an increased Alfv\'en radius, which enables an easier crossing of the Alfv\'enic transition. Inside an LMBL, the amplitudes of SBs are suppressed by the low Alfv\'en Mach number. A probable source on the Sun for an LMBL is the peripheral region inside a coronal hole with rapidly diverging open fields, so an LMBL can be considered as a transition layer between slow and fast wind. The typical angular width of an LMBL is expected to be a few degrees on the Sun, but its timescale may vary depending on how long the magnetic footpoint of PSP lingers at the source. 

(2) The sub-Alfv\'enic wind detected by PSP is a special type of wind emanating from a coronal hole near its boundary along rapidly diverging open fields, i.e., an LMBL flow by nature. It is observable even at a distance of about 20 solar radii from the Sun because of its enhanced Alfv\'en radius. The height of $\sim$20 solar radii should not be regarded as the typical Alfv\'en radius. The average value is 10 - 12 solar radii. In the sub-Alfv\'enic wind, Alfv\'enic deflections do not disappear, although they are reduced by the low Alfv\'en Mach number. A preliminary look at some other sub-Alfv\'enic intervals from the latest PSP encounters seems to confirm the typical characteristics identified here (e.g., decreased Alfv\'en Mach number, reduced density, relatively low velocity, and suppressed SBs).

(3) The LMBLs including the sub-Alfv\'enic wind all have a similar origin (i.e., coronal hole boundaries with rapidly diverging open fields) and similar properties (i.e., reduced density, relatively low velocity, and signatures resembling fast wind). These favor a wrinkled surface for the morphology of the Alfv\'enic transition, rather than an extended, fragmented zone with mixed parcels of sub-Alfv\'enic and super-Alfv\'enic wind. In the context of our results, the picture of a ``rugged" Alfv\'en surface predicts higher frequency of sub-Alfv\'enic intervals as PSP descends to lower perihelia. 

(4) We find a dependence of the amplitudes of SBs on the radial Alfv\'en Mach number. The distribution, which resembles a ``herringbone" structure, indicates that a larger deflection angle tends to be associated with a higher Mach number. The majority of deflection angles are below $90^{\circ}$, so most magnetic field lines are not deflected backward but just ``sideways". We suggest that the term of SBs is better changed to Alfv\'enic flows with deflected magnetic field and enhanced radial velocity, or Alfv\'enic deflections for short. The magnetic deflections have an origin well below the Alfv\'en critical point (i.e., $M_{\rm A}<1$), and deflection angles larger than $90^{\circ}$ seem to occur only when $M_{\rm A} \gtrsim 2$, i.e., well above the Alfv\'en critical point. 

(5) We present a new approach to calculating the deflection angle, which indicates the deflection direction and allows to establish a relationship with the enhancement in the solar wind radial velocity. A preferential clockwise deflection is observed in the RT plane as viewed from the north. This is attributed to the Parker spiral geometry of the mean magnetic field, or the rotation of the Sun.     

(6) A simple analytical model is developed for the relationship between the deflection angle and velocity variation. As indicated by the model, the velocity change is always positive for any deflection of the field, which explains the one-sided nature of the velocity spikes observed in SBs. It suggests that below the Alfv\'en critical point the velocity enhancement and the deflection angle must be significantly reduced. This agrees with the observations in the LMBLs including the sub-Alfv\'enic wind. The model also predicts that obvious SB remnants at 1 AU and beyond may correspond to a high Alfv\'en Mach number and a background field geometry with a large radial component. These are simply consequences of outward propagating Alfv\'enic fluctuations.   

(7) The velocity enhancement in units of the local Alfv\'en speed generally increases with the deflection angle. This is consistent with our analytical model. A surprising finding is that, while there are quite some deflection angles larger than $90^{\circ}$, the data points with $\delta{v_R} > v_{\rm A}$ are very few. This finding likely indicates a nonlinearly evolved, saturated state where the local Alfv\'en speed is roughly an upper bound for the velocity enhancement. 

(8) We test the theories on the origin of SBs using our results about SBs and the LMBLs. Among the theories (i.e., interchange reconnection, velocity shear and footpoint motion, expanding waves and turbulence, and coronal jets), the most promising one in the context of our work is the model of expanding waves and turbulence. Others do not quite fit our results. Note that the theory of expanding waves and turbulence should not be considered as only an in situ formation process in the solar wind. We find that SBs have an origin well below the Alfv\'en critical point. 

(9) The results about the LMBLs and dependence of SBs on the Alfv\'en Mach number indicate that the patchy distribution of SBs is due to modulation by reductions in the Mach number. In general, a correlation is observed between the drop in the Mach number and the reduction of SBs. A repeated in-and-out motion of the PSP footpoint with respect to one or more LMBLs can readily explain the patchy distribution, no matter whether it is a spatial or temporal effect.

(10) Finally, we create a picture on the generation and evolution of SBs based on our results. Photospheric motions shake magnetic field lines and drive Alfv\'en waves that propagate outward \citep[][and references therein]{cranmer2009}. As the solar plasma expands and accelerates to become the solar wind, the outward propagating Alfv\'en waves are able to produce deflections in the magnetic field and enhancements in the plasma radial velocity. Although weak below the Alfv\'en critical point, their amplitudes develop when the Alfv\'en Mach number increases. At a certain Mach number or distance well above the Alfv\'en critical point, the amplitudes are such that the magnetic field begins to be deflected backward (i.e., true SBs form). Deflections larger than $90^{\circ}$ represent a well evolved state, where the local Alfv\'en speed can be regarded as an upper bound for the velocity enhancement. At a certain stage, a decay process may come into play \citep[e.g.,][]{tenerani2021}. SBs begin to fade before reaching 1 AU. Clear remnants observed at 1 AU or beyond correspond to only certain circumstances. 

\acknowledgments The research was supported by NSFC under grants 42274201 and 42004145, and the Specialized Research Fund for State Key Laboratories of China. We acknowledge the NASA Parker Solar Probe mission and the SWEAP and FIELDS teams for use of data. The PFSS extrapolation is performed using the \emph{pfsspy} Python package \citep{stansby2020}. The data used for PFSS modeling are courtesy of GONG and SDO/AIA.

\clearpage

\begin{figure}
\epsscale{0.9} \plotone{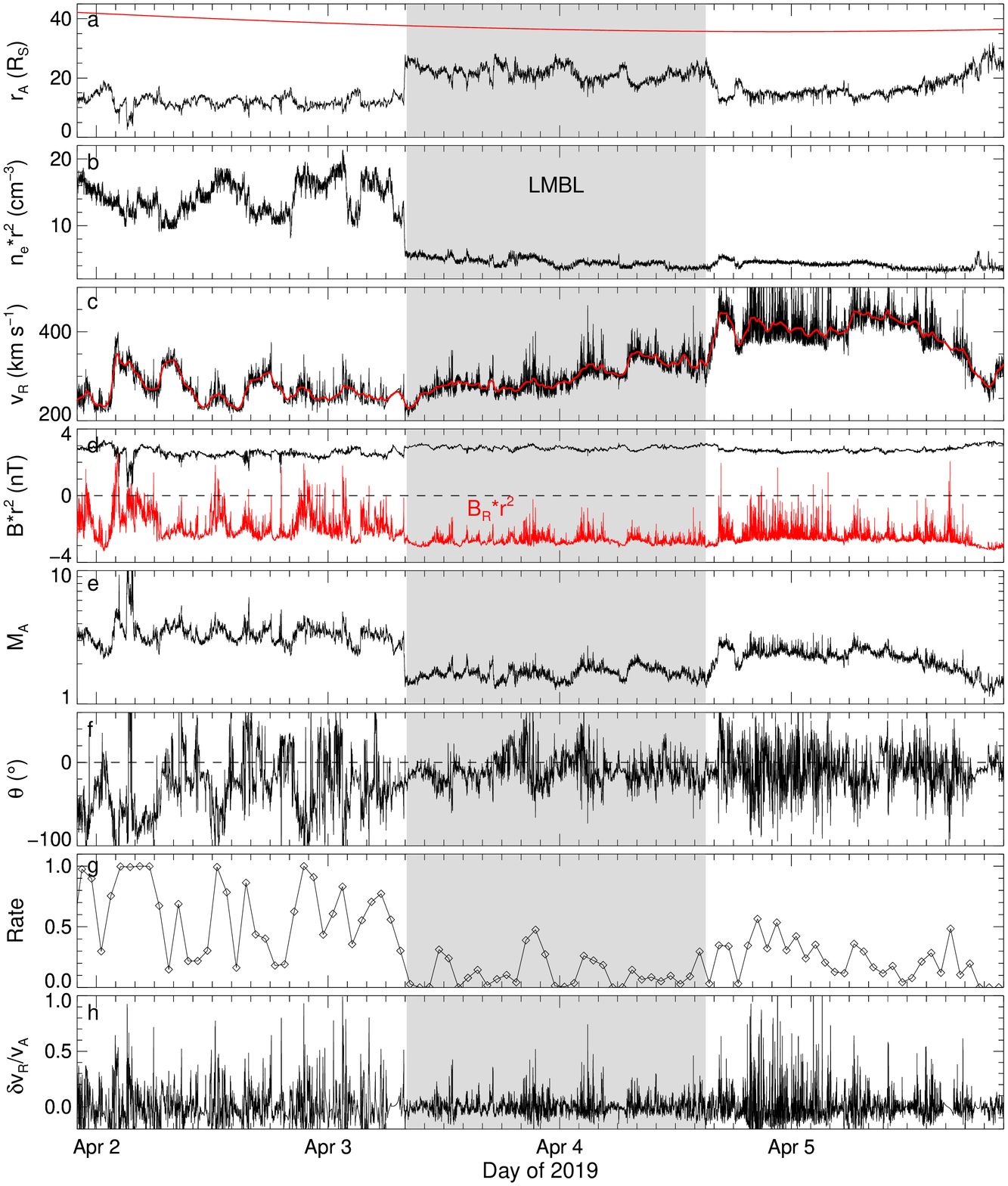} 
\caption{PSP measurements at the second encounter. The shaded region indicates the LMBL interval. (a) Alfv\'en radius in comparison with the distance of the spacecraft (red). (b) Electron density (normalized to 1 AU values) from quasi-thermal noise (QTN). (c) Proton radial velocity and filtered values (red). (d) Normalized magnetic field strength and radial component (red). (e) Radial Alfv\'en Mach number. (f) Magnetic field deflection angle. (g) Hourly occurrence rate of $|\theta| \geqslant 30^{\circ}$. (h) Radial velocity enhancement in units of local Alfv\'en speed.} 
\end{figure}

\clearpage

\begin{figure}
\epsscale{0.9} \plotone{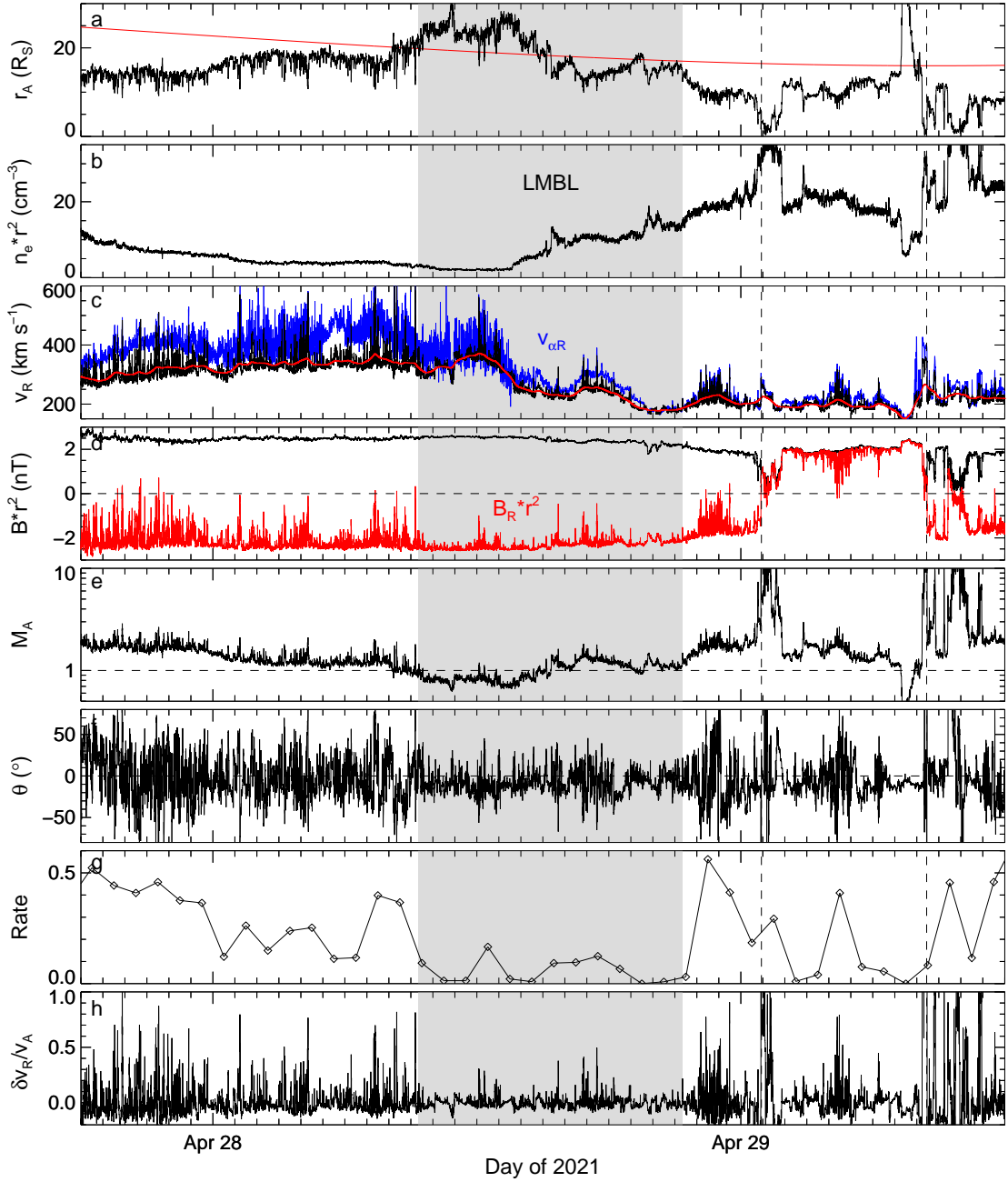} 
\caption{PSP measurements at the eighth encounter. Similar to Figure~1. Also shown in panel (c) is the radial velocity of alpha particles (blue). The vertical dashed lines mark HCS crossings.}
\end{figure}

\clearpage

\begin{figure}
\epsscale{0.9} \plotone{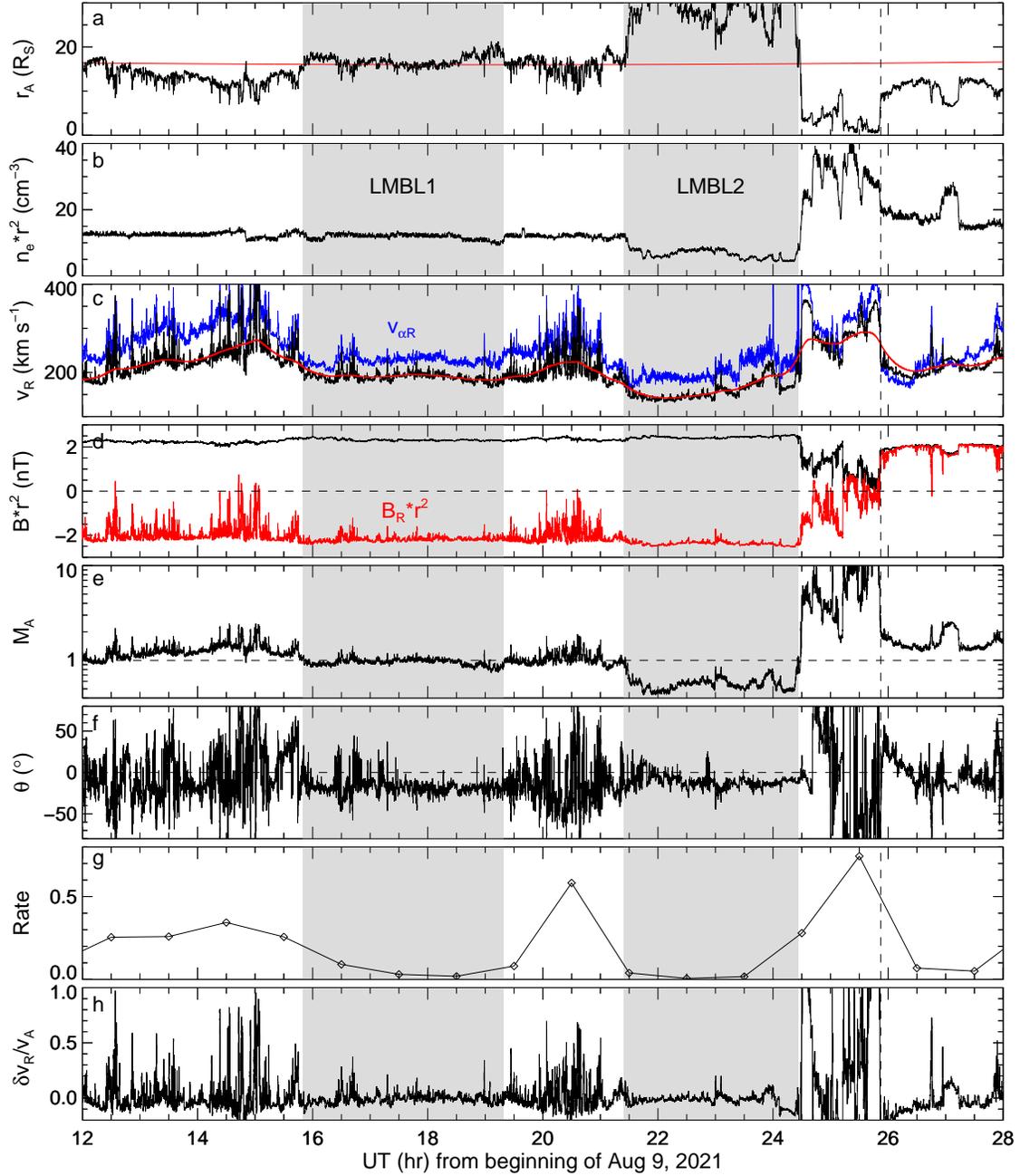} 
\caption{PSP measurements at the ninth encounter. Similar to Figures 1 and 2. Here two LMBLs are identified.}
\end{figure}

\clearpage

\begin{figure}
\epsscale{1.0} \plotone{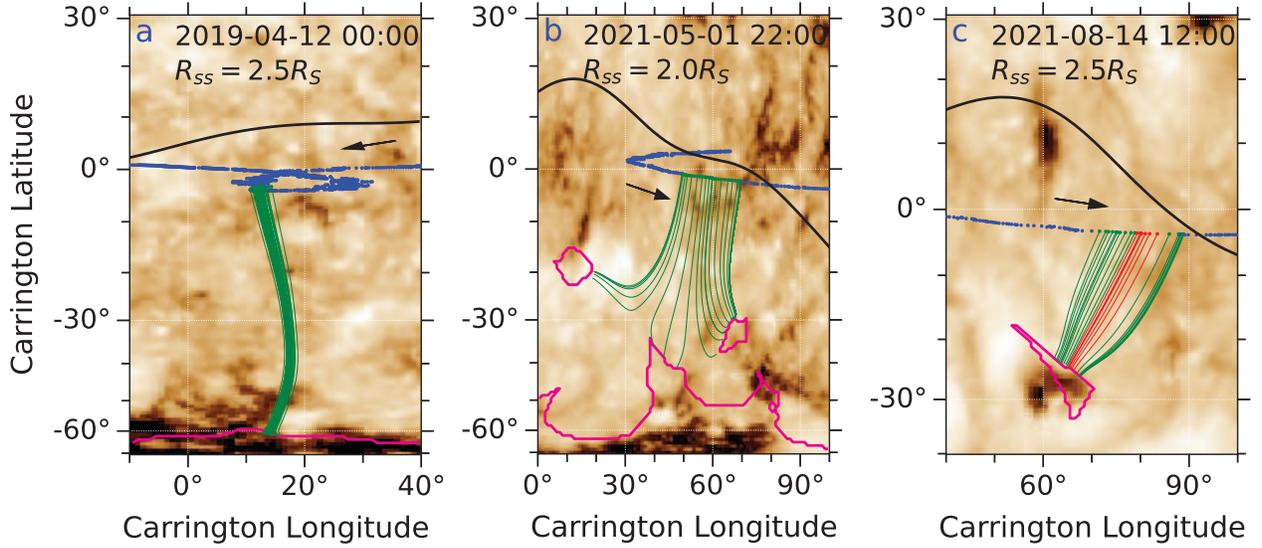} 
\caption{Magnetic mapping of the LMBLs back to the solar surface. (a) Encounter 2. (b) Encounter 8. (c) Encounter 9. An EUV synoptic map from SDO observations at 193 \AA\ is used as a background. The pink contours outline the areas of the coronal holes of interest from PFSS modeling, and the black line marks the source surface neutral line (the coronal base of the heliospheric current sheet). The blue curve is the trajectory of PSP projected onto the source surface, and the black arrow shows the direction of motion of the spacecraft. The green lines indicate the connectivity of the LMBLs from the PSP trajectory to the photospheric sources. The red lines in panel (c) represent the connectivity for the interval between the two LMBLs in Figure~3. Also shown are the time tag of the ADAPT-GONG magnetogram and the height of the source surface used in the PFSS modeling on the top of each panel.}
\end{figure}

\clearpage

\begin{figure}
\epsscale{0.8} \plotone{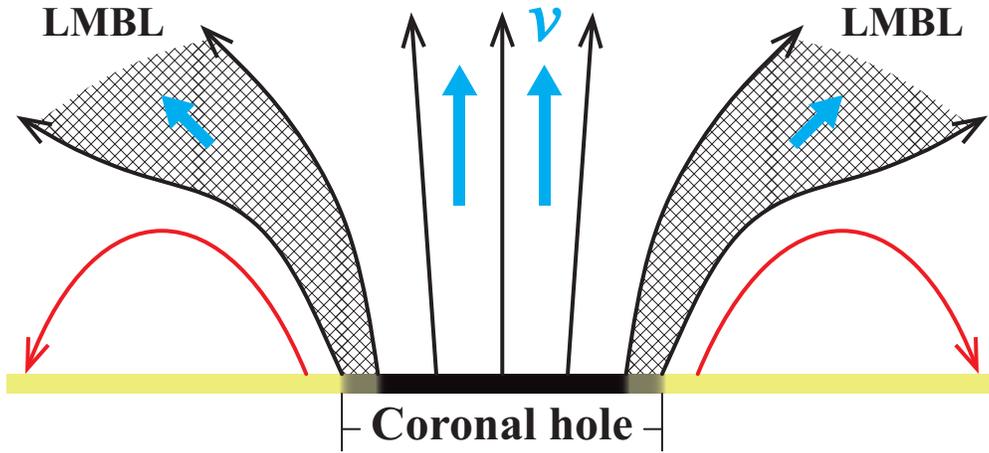} 
\caption{Schematic diagram illustrating a probable source on the Sun for an LMBL. The black lines represent open magnetic field lines from a coronal hole, and the red lines are closed field lines outside the coronal hole. The hatched area indicates the LMBL with rapidly diverging field lines rooted in the peripheral region (gray) inside the coronal hole. Deeper interior (dark) of the coronal hole is associated with slowly diverging field lines. The size of the light blue arrows shows the magnitude of the solar wind velocity.}
\end{figure}

\clearpage

\begin{figure}
\centerline{\includegraphics[width=22pc]{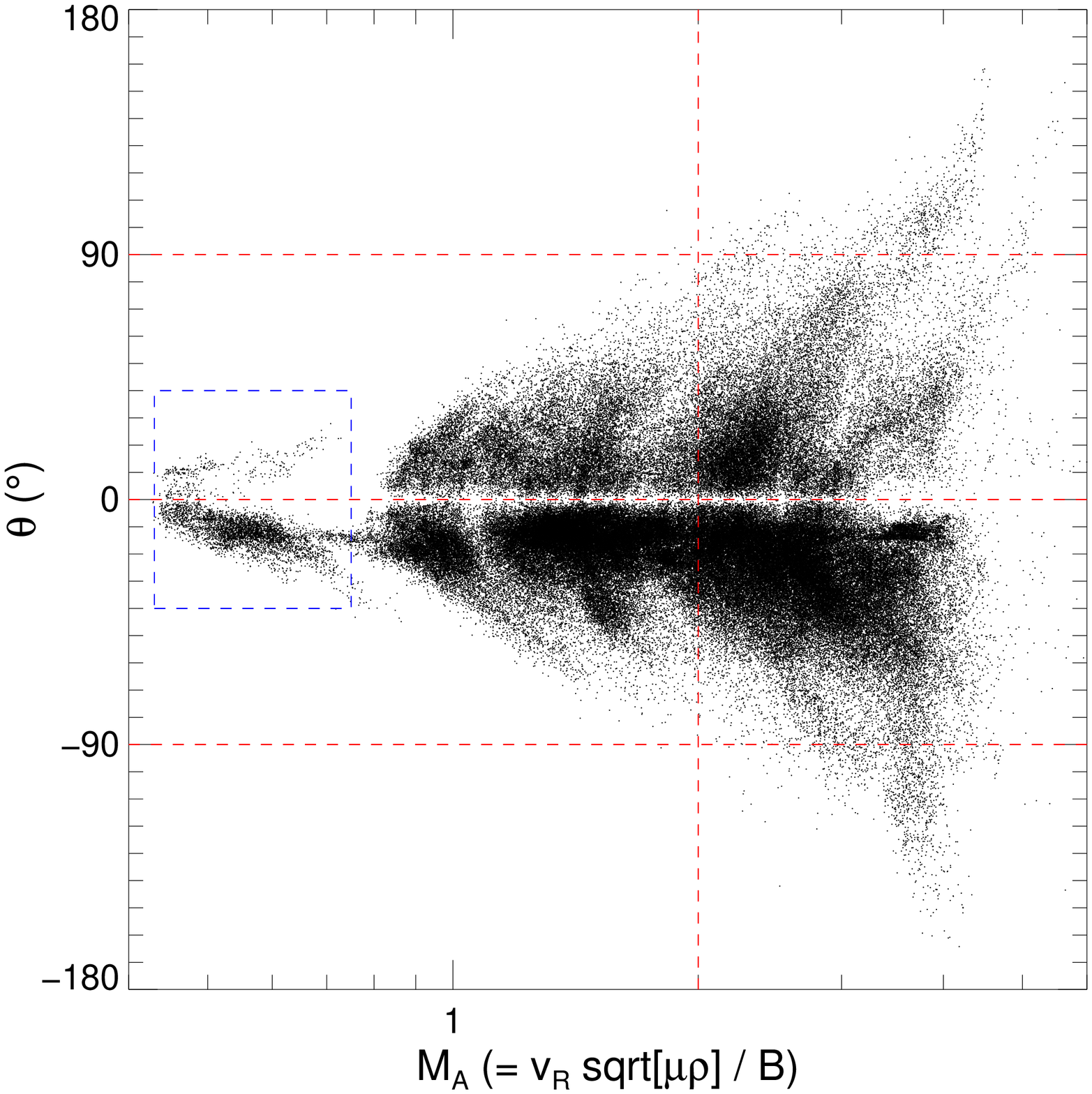} \hspace{0.1pc} \includegraphics[width=22pc]{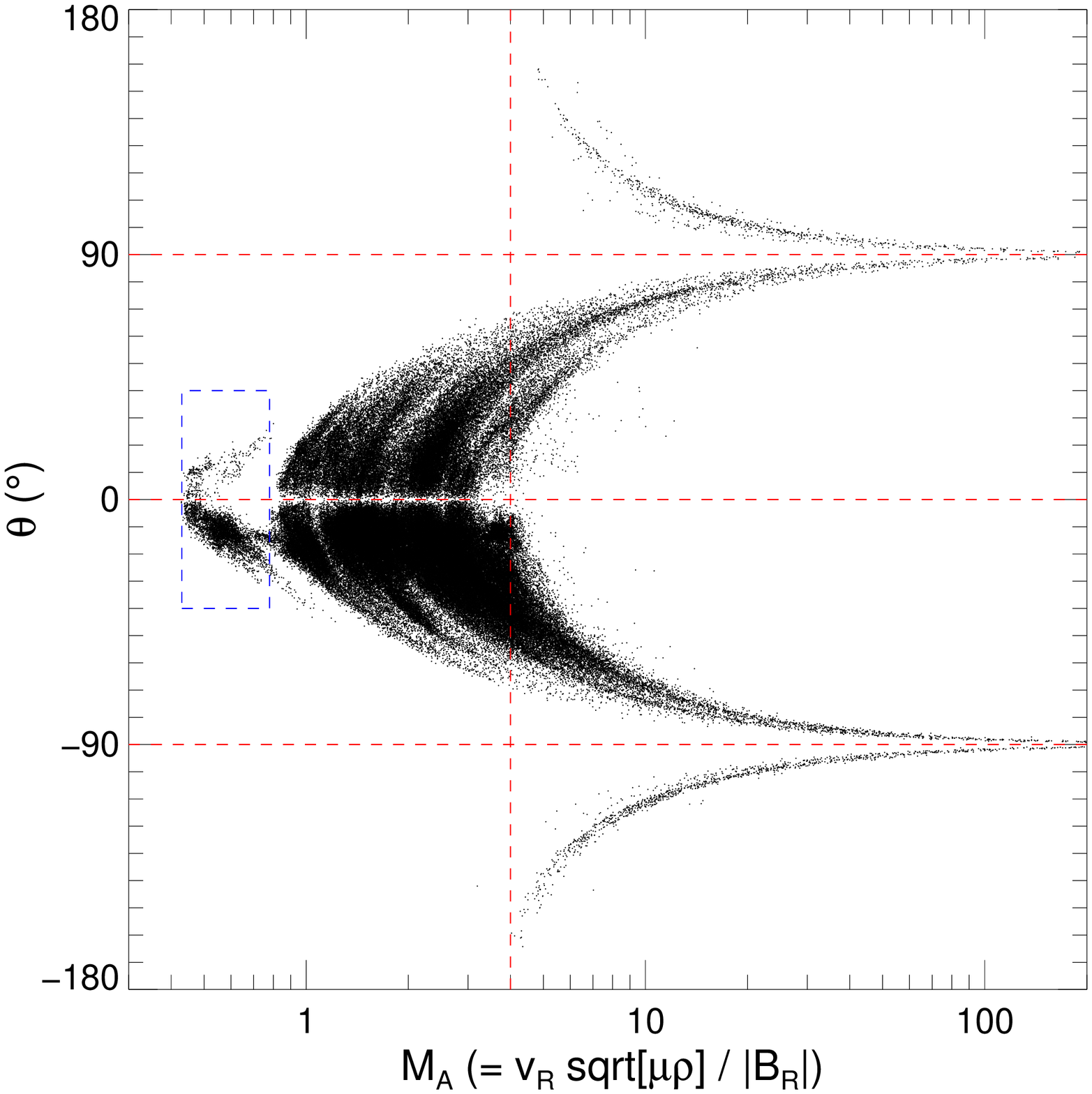}}
\caption{Encounter 9 measurements of magnetic field deflection angle as a function of radial Alfv\'en Mach number, resembling a ``herringbone" structure. Note the different definitions of $M_{\rm A}$ on the left and right panels. The dashed blue box indicates the sub-Alfv\'enic solar wind (mainly LMBL2 in Figure~3). The vertical dashed line marks the Mach number beyond which $|\theta|>90^{\circ}$ occurs.}
\end{figure}

\clearpage

\begin{figure}
\epsscale{0.9} \plotone{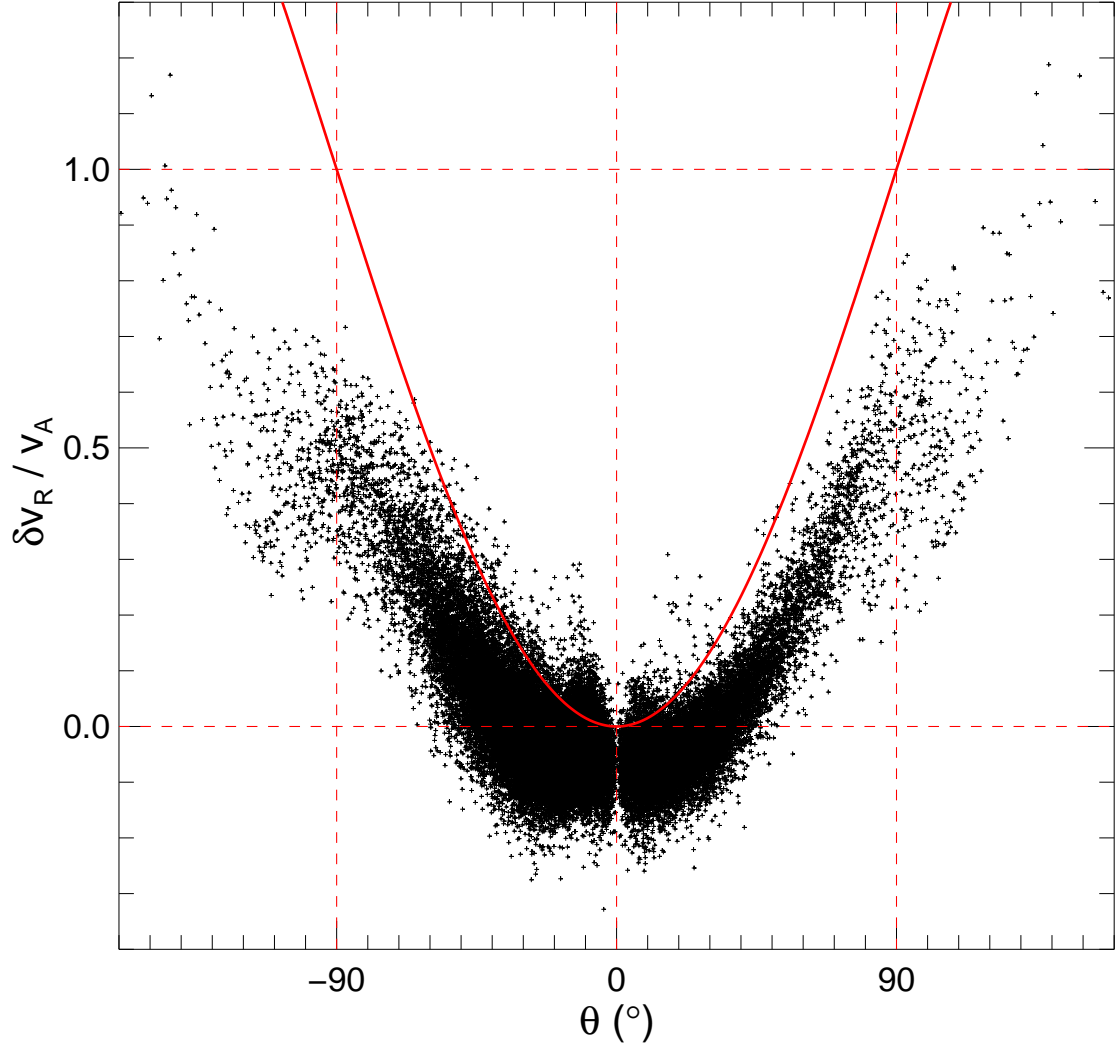} 
\caption{Encounter 9 measurements of radial velocity enhancement in units of local Alfv\'en speed as a function of magnetic field deflection angle. The red solid curve is derived from Equation~(3). There are negative values of $\delta{v_R}$ at smaller deflection angles. This happens in a filtering due to mixing of fluctuations of various frequencies.}
\end{figure}

\clearpage

\begin{figure}
\epsscale{0.9} \plotone{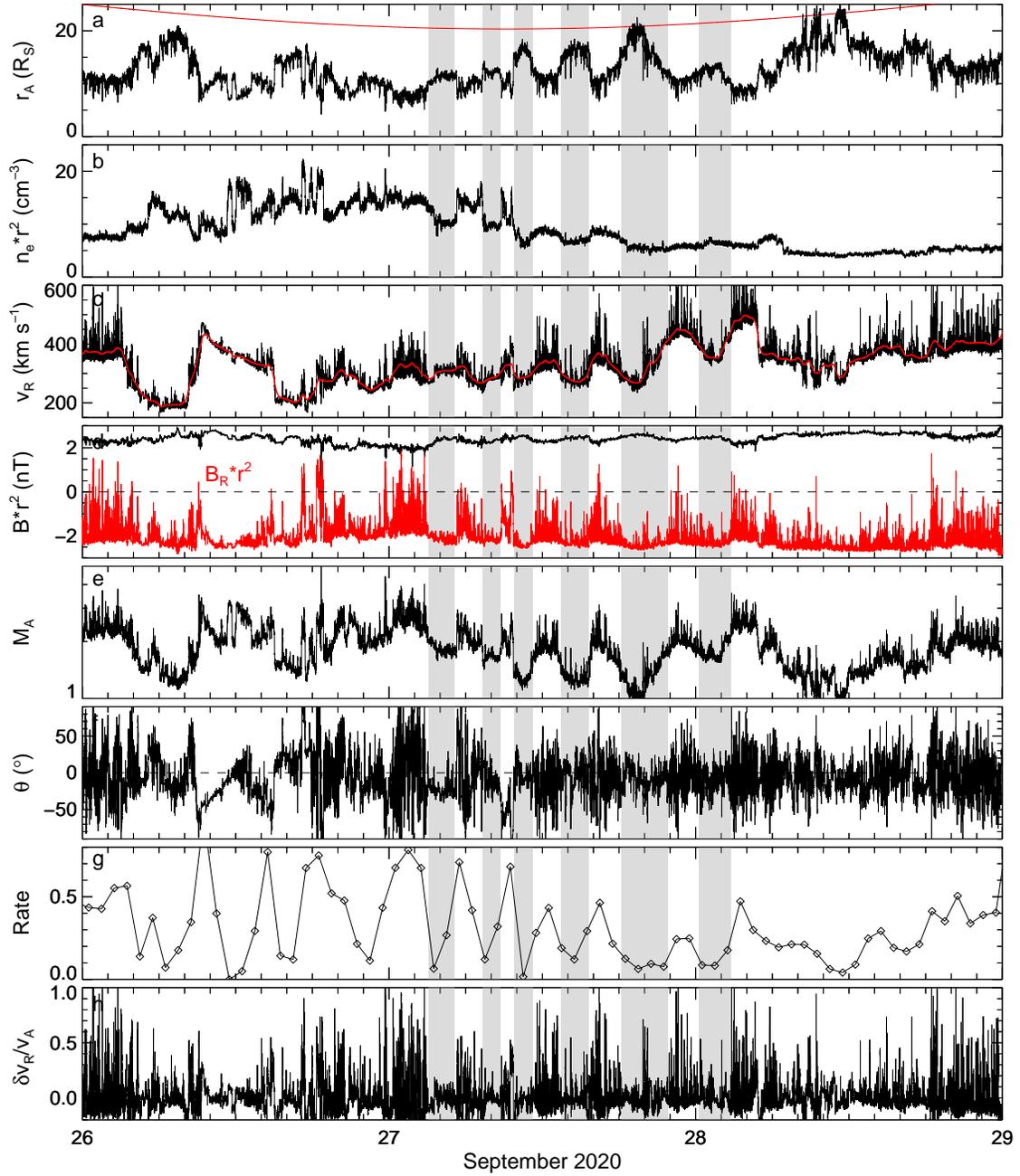} 
\caption{PSP measurements at the sixth encounter illustrating modulation of switchbacks by reduced Mach-number intervals (shaded regions). Similar to Figure~1. Note a transient structure from about 08 - 16 UT on September 26 with a coherent rotation in the magnetic field.}
\end{figure}

\end{document}